\documentclass[twocolumn,english,aps,prl,superscriptaddress,showpacs]{revtex4}
\usepackage[T1]{fontenc}
\usepackage[latin1]{inputenc}
\usepackage{graphicx}

\makeatletter

\usepackage{babel}
\makeatother
\begin{document}

\title{Pore Stability and Dynamics in Polymer Membranes}

\author{H. Bermudez}

\email{bermudez@seas.upenn.edu}

\affiliation{Department of Chemical and Biomolecular Engineering, University of
Pennsylvania, Philadelphia, Pennsylvania 19104}

\author{H. Aranda-Espinoza }

\affiliation{Institute for Medicine and Engineering, University of Pennsylvania,
Philadelphia, Pennsylvania 19104}

\author{D. A. Hammer}

\affiliation{Department of Chemical and Biomolecular Engineering, University of
Pennsylvania, Philadelphia, Pennsylvania 19104}

\affiliation{Institute for Medicine and Engineering, University of Pennsylvania,
Philadelphia, Pennsylvania 19104}

\author{D. E. Discher}

\affiliation{Department of Chemical and Biomolecular Engineering, University of
Pennsylvania, Philadelphia, Pennsylvania 19104}

\affiliation{Institute for Medicine and Engineering, University of Pennsylvania,
Philadelphia, Pennsylvania 19104}

\begin{abstract}
Vesicles self-assembled from amphiphilic diblock copolymers exhibit
a wide diversity of behavior upon electroporation, due to competitions
between edge, surface and bending energies that drive the system,
while different viscous dissipation mechanisms determine the timescales.
These effects are manifestations of the varying membrane thickness
$d$ from what are essentially chemically identical molecules. For
smaller $d$, we find large unstable pores and the resulting membrane
fragments reassemble into vesicles within minutes. Vesicles with larger
$d$ form long-lived pores of nanometer dimension that could be potentially
useful for sieving applications and drug delivery.
\end{abstract}


\maketitle
Membranes are ubiquitous in biology and play a central role in the
distribution of macromolecules both inside and outside cells. Additionally,
specialized transmembrane machinery serves to exquisitely regulate
intracellular concentrations at their optimal, non-equilibrium conditions
\cite{alberts}. These transmembrane gradients are known to be involved
in signaling and other regulatory processes. Thus the control and
exploitation of biomembrane selectivity has been and continues to
be the focus of intense research \cite{recent}. One approach to manipulate
biomembranes is known as electroporation, where an electric field
is used to transiently disrupt the membrane \cite{poration}. In the
brief time the membrane is open, solutes or macromolecules can diffuse
into or out of the cell. Technological applications for DNA delivery
and creation of hybrid cells have employed electroporation, although
cell viability is not always preserved \cite{poration}.

Necessarily generated during biological processes are transmembrane
potentials $V_{m}$ and membrane tensions $\tau $, as a result of
electrostatic or mechanical forces. The continuing obstacle to approaching
biological systems is that their complexity can obscure underlying
physical mechanisms. Taking a reductionist approach, we have chosen
to use synthetic vesicles formed from diblock copolymers, called polymersomes.
These self-assembling diblocks are amphiphilic, with the hydrophilic
portion being poly(ethylene oxide) (PEO) and the hydrophobic part
composed of either polybutadiene (PBD) or its saturated form, poly(ethylethylene)
(PEE) \cite{hillmyer}. The synthetic basis allows for creation of
a family of varying molecular weight $\bar{M}_{n}$ from $\approx 4-20$
kg/mol \cite{harry}. The resulting hydrophobic thickness $d$ ranges
from $8-21$ nm, compared to biomembranes of $3-5$ nm. Systematic
investigations of electroporation are therefore motivated from both
a biological and a physical perspective. In this Letter, we report
our observations of a wide diversity of pore behavior, owing to the
expanded range of $d$ that is available with polymeric systems.

The experimental procedure closely follows Aranda-Espinoza, \textit{et
al}. \cite{helim}. In all cases, we used the film rehydration method
to prepare vesicles, as is common for liposomes. A combination of
phase contrast and fluorescence microscopy were used to monitor any
changes in vesicle integrity that occured following pore formation
\cite{optics}.

The energetics of a circular hole of size $r$ in a flat, infinitely
thin, membrane can be most simply described by $E=2\pi r\Gamma -\pi r^{2}\Sigma $,
where $\Gamma $ and $\Sigma $ are the line and surface tensions
of the membrane, respectively \cite{simplest}. The line tension accounts
for the energetic penalty involved with having a hole and the surface
tension reflects the energy associated with a loss of membrane area.
This relation predicts a metastable pore size $r^{*}=\Gamma /\Sigma $
obtained from $dE/dr=0$. Pores of size $r<r^{*}$ will reseal and
those with $r>r^{*}$ will grow without bound. 

More complex models of pore growth include a dynamic surface tension,
predicting a stable pore size \cite{herve,wyart}, but our observations
indicate unstable pores in almost every case for \textbf{OE7} ($d=8$
nm) and \textbf{OB16} ($d=11$ nm), indicating a \textit{small} value
of $\Gamma $ relative to the surface tension $\Sigma $. However,
we need to consider the influence of membrane thickness on both $\Gamma $
and $\Sigma $. At the moment of poration, $\Sigma $ is the lysis
tension $\Sigma _{c}$, and previous work has shown that $\Sigma _{c}\sim d$
\cite{helim}. The dependence of $\Gamma $ on $d$, considering both
hydrophobic and hydrophilic pores, leads to the same scaling of $\Gamma \sim d$.
For a hydrophobic pore, the line tension is most simply the product
of the exposed length $d$ and the interfacial energy density $\gamma $:
$\Gamma \sim \gamma d.$ For a hydrophilic pore, $\Gamma $ should
scale as the bending modulus $k_{c}$ multiplied by the curavture
$\sim 2/d$. Simple elastic models predict $k_{c}\sim d^{2}$ \cite{bloom}
and hence $\Gamma \sim d$ again. 

We therefore have the counterintuitive result that the metastable
pore size does not depend on membrane thickness. We do indeed see
large pores from \textbf{OE7} to \textbf{OB16} but \textbf{OB18} (\textbf{$d=15$}
nm) and \textbf{OB19} ($d=21$ nm) exhibit significantly smaller,
and long-lived, pores. This abrupt change is presumed to be a result
of increasing viscosity or chain entanglements within the hydrophobic
region of the membrane \cite{harry,lee}. The lifetime of pores is
largely dictated by the dynamics of the membrane. Liposomes are known
to reseal quickly, with a few notable exceptions \cite{stable}. Under
no external stresses, viscous dissipation at the interface and within
the membrane are the obstacles to in-plane material rearrangement.

\begin{figure}[!htbp]
\begin{center}\includegraphics[  width=3.25in,
  keepaspectratio]{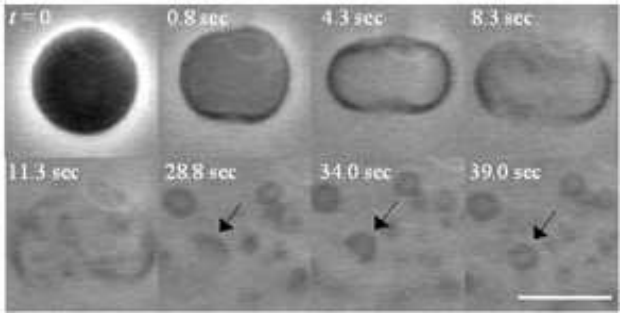}\end{center}

\caption{Time sequences of an \textbf{OE7} ($d=8$ nm) vesicle following electroporation.
The first reassembled vesicle appears at $t\approx 26$ sec, indicative
of a large edge energy compared to the bending energy. The arrows
indicate a particular disk reassembling. Scale bar is 10 $\mu $m.\label{fig:OE7}}
\end{figure}

For large pores (> 1 $\mu $m) such as those in \textbf{OE7} and \textbf{OB16},
we can visualize $r(t)$ by optical microscopy. Although both \textbf{OE7}
and \textbf{OB16} exhibit unstable pores, they both reassemble rather
quickly (Fig. \ref{fig:OE7}). To our knowledge this is the first
direct visualization of vesicle reassembly. Prior work has been focused
in the suboptical regime and timescales for assembly have been of
order minutes \cite{reseal}. This closure reflects the relative large
magnitude of the edge energy (line tension) compared to the bending
energy, characterized by a {}``vesiculation index'' $V_{f}\sim \Gamma R_{d}/k_{c}$
\cite{fromherz}, where $R_{d}$ is the radius of a disk and $k_{c}$
the bending modulus. By assuming an attempt frequency $1/\tau _{z}$
based on viscous drag in the solution \cite{cates}, we can calculate
for \textbf{OE7}, $V_{f}=2\left[1-\sqrt{k_{b}T\ln (\tau _{a}/\tau _{z})/8\pi k_{c}}\right]\approx 1.92$.
This reflects the rapid transition of \textbf{OE7} into vesicles (\textit{i.e}.,
small reassembly time $\tau _{a}$) from what are presumably disk-like
or cylindrical micelles.

\begin{figure}[!htbp]
\begin{center}\includegraphics[  width=3.25in,
  keepaspectratio]{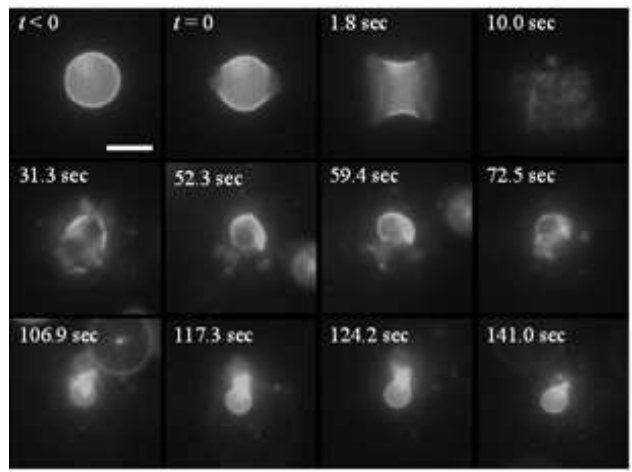}\end{center}

\caption{Time sequences of \textbf{OB16} ($d=11$ nm) following poration are
visualized with fluorescently labeled membranes. Vesicle reassembles
at $t\approx 110$ sec. Note disk-like intermediates with rough edges.
Scale bar is 10 $\mu $m.\label{fig:OB16}}
\end{figure}

\begin{figure}[!htbp]
\begin{center}\includegraphics[  width=3in,
  keepaspectratio]{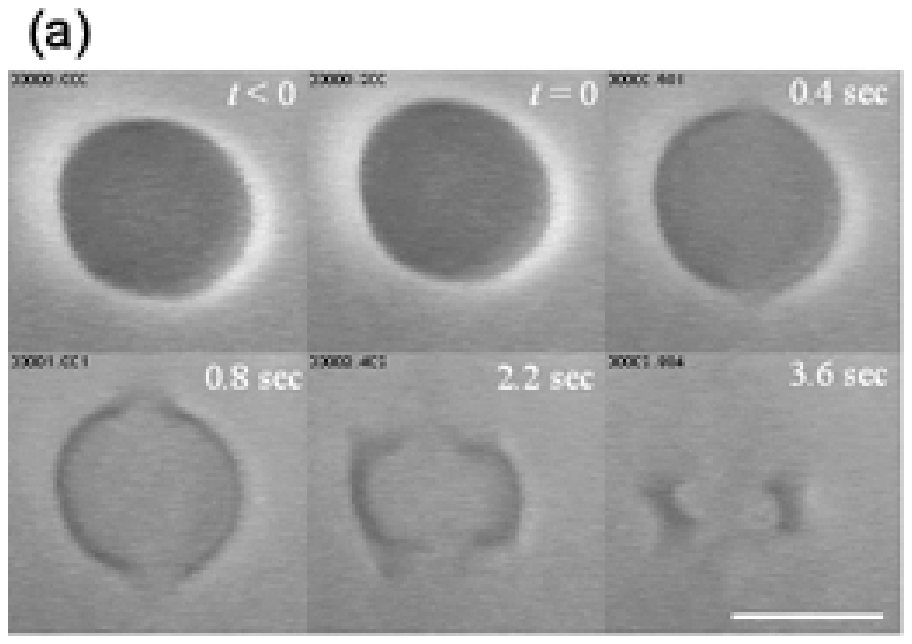}\end{center}

\begin{center}\includegraphics[  width=3.25in]{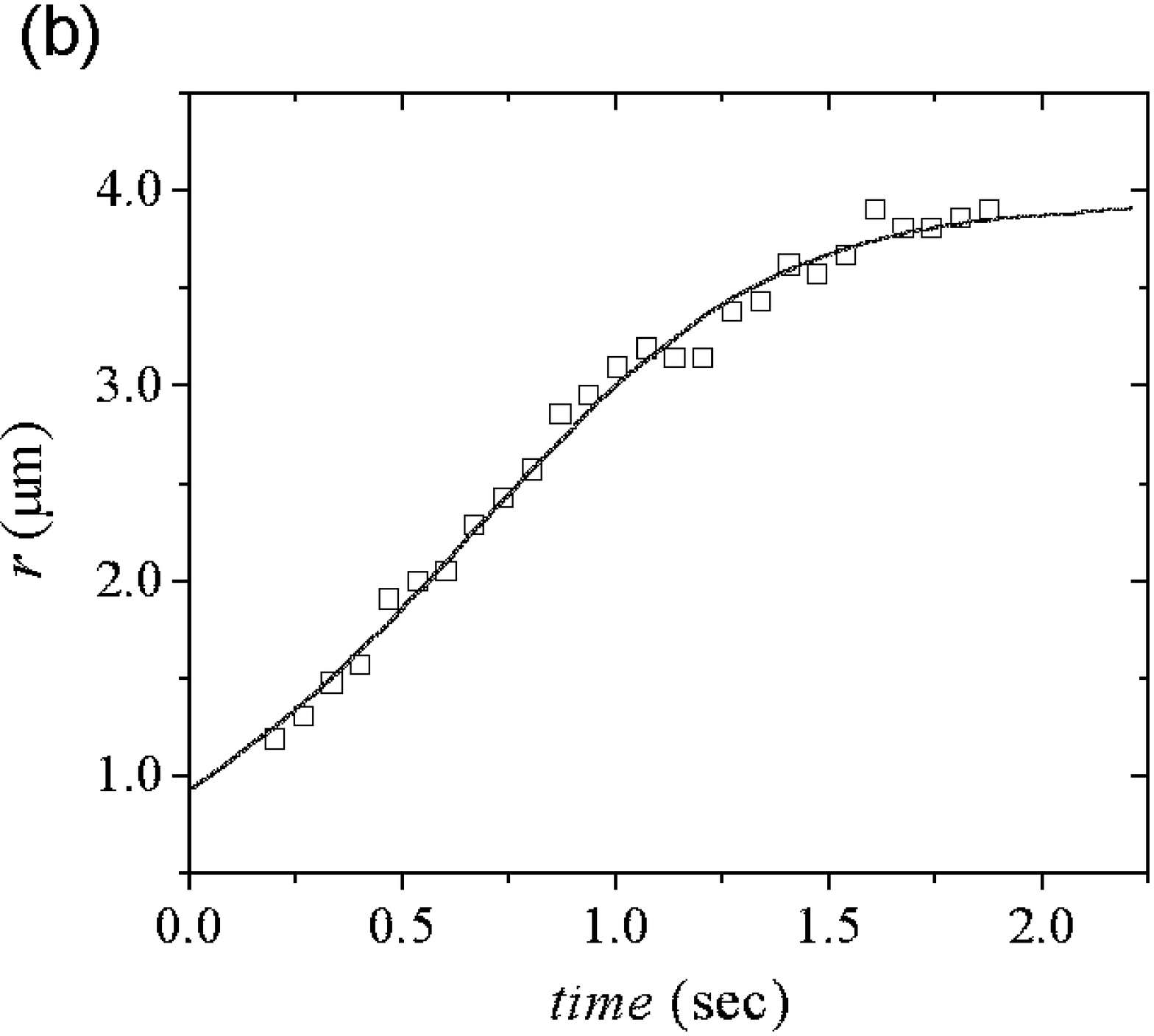}\end{center}

\caption{(a) Phase constrast imaging and (b) corresponding plot of pore radius
$r$ growth with time for \textbf{OB16}. The line is a theoretical
fit that is dominated by viscous losses within the membrane. The resulting
time constant $\tau _{1}$ is $0.67\pm 0.05$ sec ($N=6$). The error
in measuring $r$ is $\pm 0.5\, \mu $m, limited by optical resolution.
Scale bar is 10 $\mu $m\label{fig:growth}}
\end{figure}

Membranes of \textbf{OB16} reassemble over a few minutes, indicating
a smaller value of $V_{f}$ compared to \textbf{OE7}. Visualization
of the \textbf{OB16} process was possible only with fluorescently
labeled membranes, as the fragments drift substantially before reassembling.
Initially, membrane is clearly lost as the pores grow to several microns
in diameter (Fig. \ref{fig:OB16}), again suggestive of a small line
tension relative to the surface tension. The hydrophobic dye indicates
that \textbf{OB16} has an increased preference to form cylindrical
or disk-like aggregates upon loss of membrane integrity (Fig. \ref{fig:OB16}).
This is consistent with observations that morphological boundaries
(\textit{e.g}., lamellar-to-cylindrical) shift toward lower hydrophilic
block fraction with increasing $\bar{M}_{n}$ \cite{won}, as well
as with the longer reassembly time. \textbf{OE7} and \textbf{OB16}
have characteristic reassembly times $\tau _{a}$ that are 20 and
100 sec, respectively. From scaling arguments \cite{cates}, one can
show that $\tau _{a}\sim \exp (d^{2})$, indicating that even if a
larger copolymer such as \textbf{OB19} were able to reassemble, the
timescale would be several hours. Previous investigations of egg lecithin
vesicle reassembly could not be visualized due to the rapid transition
from disks to vesicles \cite{from-cryo}. Only with the addition of
an {}``edge-active'' agent could disks be made sufficiently metastable
for observation via electron microscopy. These findings are not surprising
given that using the scaling relation above and taking $d=3$ nm for
a lecithin bilayer would correspond to $\tau _{a}$ of only a few
seconds.

By extending the analysis of hole growth in thin polymer films \cite{debregeas},
several groups have developed a framework for describing pore growth
in cells and artificial vesicles \cite{herve,wyart}. There is an
initial, viscous dissipation within the membrane which at a later
time crosses over to a regime dominated by dissipation in the surrounding
fluid. For \textbf{OB16} (Fig. \ref{fig:growth}a), the only system
we can unambiguously observe, we find that the pore growth process
is almost entirely described by viscous losses within the membrane.
The time constant for the initial regime $\tau _{1}$ ($0.67\pm 0.05$
sec) allows us to determine the membrane viscosity $\eta _{m}=\tau _{1}\Sigma /d\approx 10^{6}$
Pa sec. The resulting viscosity is three orders of magnitude larger
than typical lipid membrane viscosities calculated by pore growth
analysis ($\eta _{m}\approx 10^{3}$ Pa sec \cite{wyart}), consistent
with findings by other groups using different techniques \cite{dimova}.
In the limit where the dissipation is primarily due to the surrounding
fluid viscosity $\eta _{i}$, pores will grow at a constant velocity
\cite{joanny}. From Fig. \ref{fig:growth}b, this is clearly not
the case. We can also calculate the crossover radius $r_{c}$ to the
inertial regime as $\eta _{m}d/\eta _{i}\sim 10$ meters -- indicating
that this second regime is never reached for our micron-size vesicles. 

\begin{figure}[!htbp]
\begin{center}\includegraphics[  width=2.2in,
  keepaspectratio]{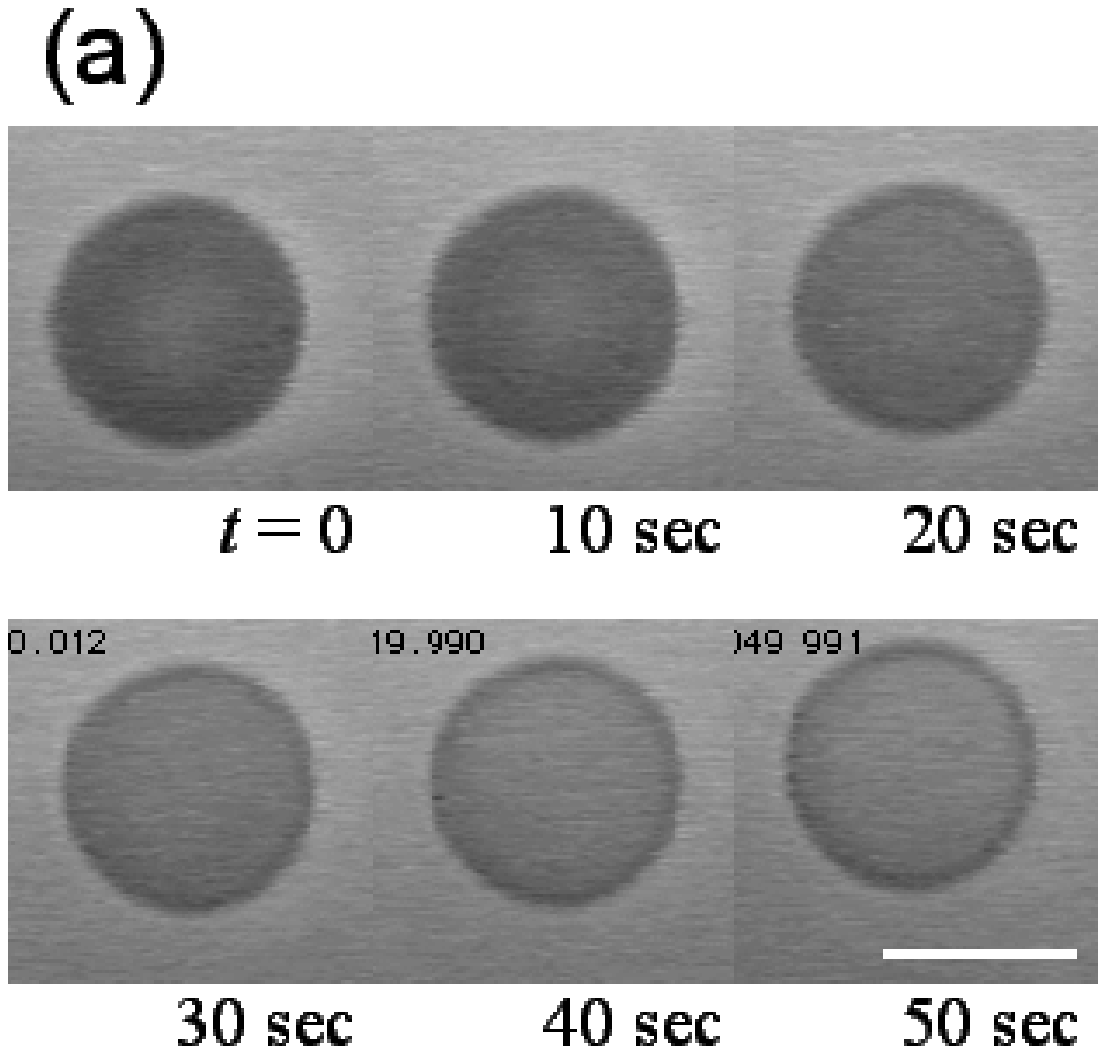}\end{center}

\begin{center}\includegraphics[  width=3.25in]{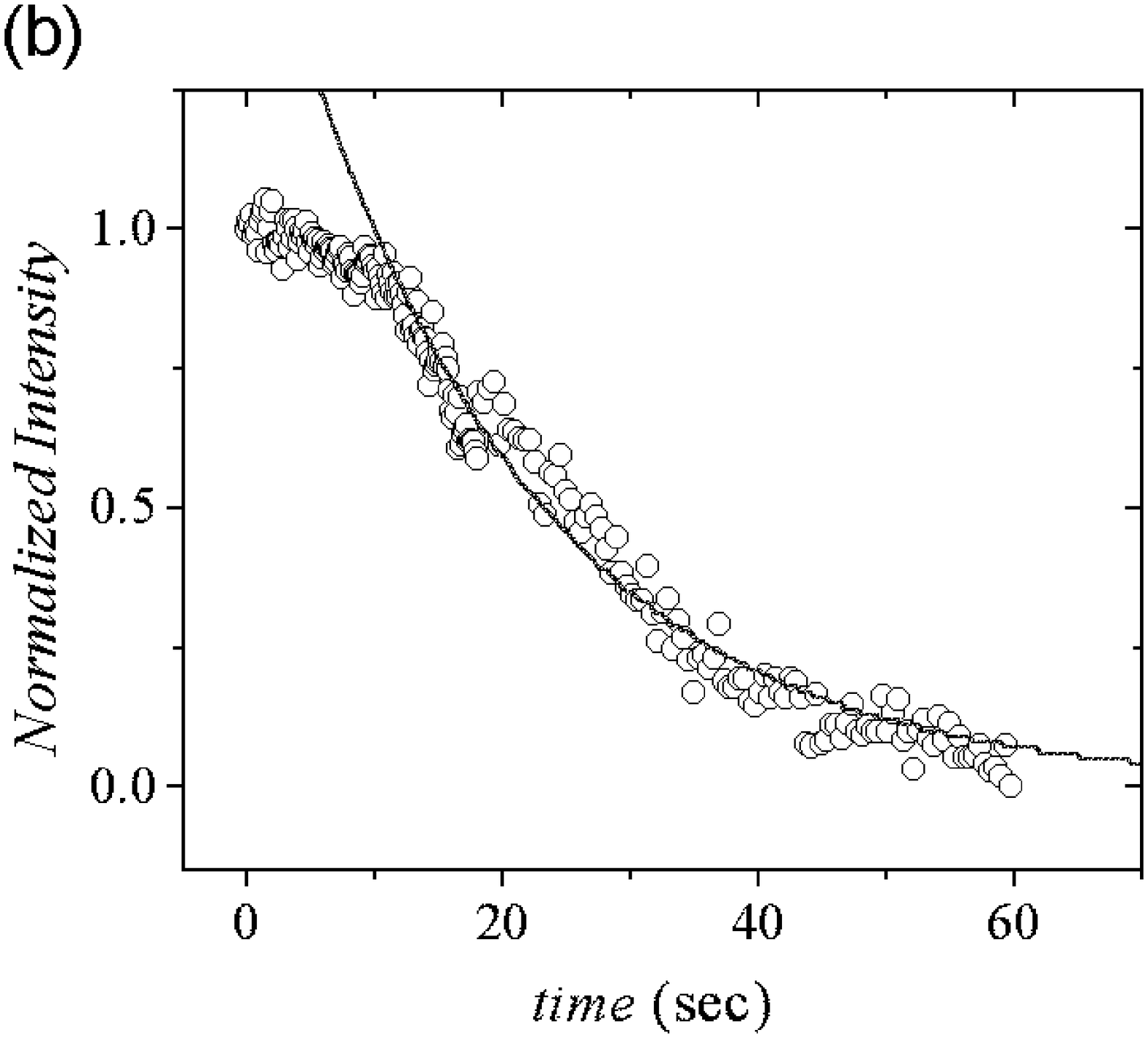}\end{center}

\caption{(a) Loss of phase contrast in \textbf{OB19} ($d=21$ nm) due to escape
of sucrose. (b) Corresponding normalized intensity of encapsulated
sucrose versus time. Solid line is a fit to diffusion through a porous
membrane, giving an effective permeability $\omega =0.10\pm 0.03$
$\mu $m/sec ($N=3$). Scale bar is 10 $\mu $m.\label{fig:OB19}}
\end{figure}
\begin{figure*}[!htbp]
\includegraphics[  width=6in,
  keepaspectratio]{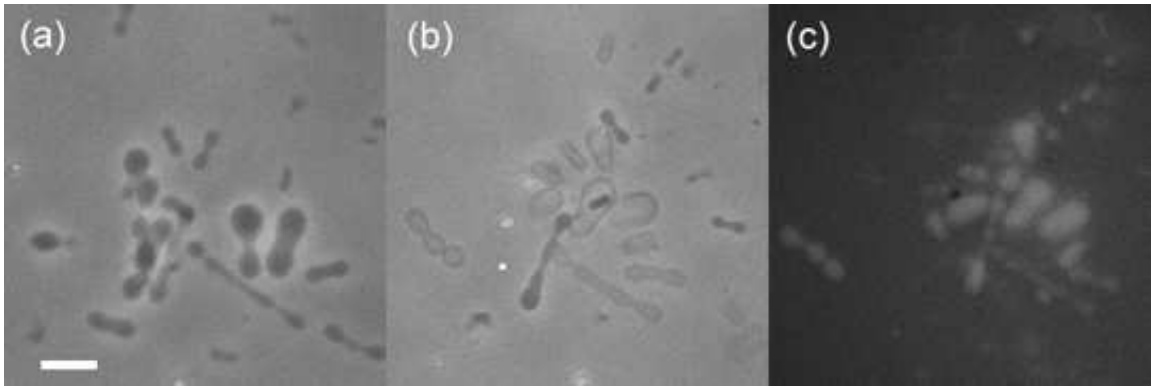}

\caption{\textbf{OB18} electroporation and encapsulant retention. (a) $t<0$,
sucrose appears dark under phase contrast microscopy. (b) At $t=1.2$
min, phase contrast indicates that sucrose has escaped, while (c)
fluorescence shows that TRITC-dextran (160 kg/mol) is still retained.
Scale bar is 25 $\mu $m.\label{fig:dextran}}
\end{figure*}

The small pores (< 1 $\mu $m) in \textbf{OB18} and \textbf{OB19}
require a slightly different analysis. In this case, the intensity
of encapsulant $I(t)$ is monitored, as pores cannot be directly observed
(Fig. \ref{fig:OB19}a). For a vesicle of size $R_{v}$, there is
an initial ``leak-out'' regime, where solute loss is driven by Laplace
pressure $\Delta P=2\Sigma /R_{v}$, lasting several milliseconds
for nanometer-size pores. We cannot observe this regime, but the tension
will relax on this timescale $\tau _{k}\sim \eta _{i}R_{v}^{2}/r\Sigma $,
and is followed by diffusion of solute through pores \cite{wyart}.
By fitting our data (Fig. \ref{fig:OB19}b) to a simple exponential
decay model of transport through porous membranes \cite{hobbie},
we can obtain a time constant $\tau _{d}=R_{v}/3\omega $, where $\omega $
is the effective membrane permeability. There is an initial time lag
in the intensity decay $t_{lag}\approx 10$ sec, which we tentatively
attribute to hindered release of solute due to the PEO chains lining
the nascent pores. The time constant increases with membrane thickness
($\tau _{d}=13\pm 4$ sec for \textbf{OB18} and \textbf{$\tau _{d}=17\pm 6$}
sec for \textbf{OB19}), reflecting either slower dynamics within the
membrane and/or smaller pore sizes. The net result is an effective
(sucrose) permeability $\omega =0.22\pm 0.15$ $\mu $m/sec for \textbf{OB18}
and $\omega =0.10\pm 0.03$ $\mu $m/sec for \textbf{OB19}. 

To estimate the size of sub-micron pores, we encapsulated fluorescent
dextrans of various molecular weights and subjected these vesicles
to electroporation (Fig. \ref{fig:dextran}). Sucrose, which can be
viewed under phase contrast and has a size $\approx 0.6$ nm, always
escapes the porated membrane. The FITC-labeled dextran of $\bar{M}_{n}\approx 4.4$
kg/mol and a TRITC-labeled dextran of $\bar{M}_{n}\approx 160$ kg/mol
were retained over approximately 10 minutes. The sizes of these dextrans
can be estimated by scaling relations \cite{hobbie} to give hydrodynamic
radii of $R_{h}\approx 2$ nm and $R_{h}\approx 5$ nm for the FITC
and TRITC-dextrans, respectively. Thus although the pore size is only
a few nanometers, given the dextran polydispersity ($\approx 1.5$),
more precise measurements on the pore size will be published elsewhere
\cite{xxx}.

In summary, we have observed for the first time vesicle formation
from disk-like fragments, allowing us to obtain scaling relations
for reassembly time. In certain cases, the slow dynamics of pore growth
can be used to obtain the membrane viscosity (\textit{e.g.}, \textbf{OB16}
$\eta _{m}=10^{6}$ Pa sec), qualitatively consistent with observations
by other groups \cite{dimova}. Sucrose leakage from thicker membranes
(\textbf{OB18} and \textbf{OB19}) gives effective membrane permeabilities
of vesicles with nanometer-scale pores. Control of the size and number
of these pores by methods such as crosslinking \cite{xlink} invites
speculation on possible applications such as controlled drug delivery
as well as for experimental tests of predictions for (bio)polymer
translocation through pores \cite{dna-pore,dna-theory}. The results
illustrate the rich diversity of behavior possible by changing a critical
length scale $d$, with other chemistries potentially providing even
more distinctive effects.

\begin{acknowledgments}
Funding was provided NSF-MRSEC's at Penn and the University of Minnesota,
and NASA. The authors thank P. Photos for stimulating discussions. 
\end{acknowledgments}


\begin{thebibliography}{10}
\bibitem{alberts}B. Alberts, \textit{et al.}, \textit{Essential Cell Biology} (Garland,
New York, 1997).
\bibitem{recent}P. Barton, \textit{et al.}, Angew. Chem. Int. Ed. \textbf{41}, 3878
(2002).
\bibitem{poration}E. Neumann, \textit{et al.}, \textit{Electroporation and Electrofusion
in Cell Biology} (Plenum, New York, 1989).
\bibitem{hillmyer}M.A. Hillmyer and F.S. Bates, Macromolecules \textbf{29}, 6994 (1996).
\bibitem{harry}H. Bermudez, \textit{et al.}, Macromolecules \textbf{35}, 8203 (2002).
\bibitem{helim}H. Aranda-Espinoza, \textit{et al.}, Phys. Rev. Lett. \textbf{87},
208301 (2001).
\bibitem{optics}We used 0.3 M sucrose as our rehydrating solution, with vesicles later
suspended in an osmotically-matched glucose solution, thus settling
to the bottom of the chamber. The density difference coincides with
a refractive index difference, allowing the use of phase contrast
microscopy. Fluorescent labeling of the membrane is accomplished by
incorporation of a hydrophobic dye (PKH26, Sigma) into the bilayer,
which permits visualization of any membrane fragments or cylindrical
micelles that might be present.
\bibitem{simplest}J.D. Litster, Phys. Lett. \textbf{53A}, 193 (1974); B.V. Derjaguin
and Yu.V. Gutop, Kolloid Zh. \textbf{24}, 431 (1962).
\bibitem{herve}H. Isambert, Phys. Rev. Lett. \textbf{80}, 3404 (1998)
\bibitem{wyart}O. Sandre, L. Moreaux, F. Brochard-Wyart, Proc. Natl. Acad. Sci. USA
\textbf{96}, 10591 (1999); F. Brochard-Wyart, P. G. de Gennes, O.
Sandre, Physica A \textbf{278}, 32 (2000).
\bibitem{bloom}M. Bloom, E. Evans, O.G. Mouritsen, Q. Rev. Biophys. \textbf{24},
293 (1991).
\bibitem{lee}J.C-M. Lee, \textit{et al.}, Macromolecules \textbf{35}, 323 (2002).
\bibitem{stable}D. Zhelev and D. Needham, Biochim. Biophys. Acta \textbf{1147}, 89
(1993). J.D. Moroz and P. Nelson, Biophys. J. \textbf{72}, 2211 (1997).
\bibitem{reseal}Y. Xia, \textit{et al.}, Langmuir \textbf{18}, 3822 (2002).
\bibitem{fromherz}P. Fromherz, Chem. Phys. Lett. \textbf{94}, 259 (1983).
\bibitem{cates}J. Leng, S.U. Engelhaaf, M.E. Cates, Europhys. Lett. \textbf{59},
311 (2002).
\bibitem{won}Y-Y. Won, \textit{et al.}, J. Phys. Chem. B \textbf{106}, 3354 (2002).
\bibitem{from-cryo}P. Fromherz and D. R\"{u}ppel, FEBS Lett. \textbf{179}, 155 (1985).
\bibitem{debregeas}G. Debr\'{e}geas, P. Martin, F. Brochard-Wyart, Phys. Rev. Lett.
\textbf{75}, 3886 (1995).
\bibitem{dimova}R. Dimova, \textit{et al.}, Eur. Phys. J. E \textbf{7}, 241 (2002).
\bibitem{joanny}J.F. Joanny and P.G. de Gennes, Physica A \textbf{147}, 238 (1987).
\bibitem{hobbie}R.K. Hobbie, \textit{Intermediate Physics for Medicine and Biology}.
(Springer-Verlag, New York, 1997).
\bibitem{xxx}H. Bermudez, \textit{et al.} (unpublished).
\bibitem{xlink}B. Discher, \textit{et al.}, J. Phys. Chem. B \textbf{106}, 2848 (2002).
\bibitem{dna-pore}S.E. Hendrickson, \textit{et al.}, Phys. Rev. Lett. \textbf{85}, 3057
(2000); A. Meller, L. Nivon, D. Branton, Phys. Rev. Lett. \textbf{86},
3435 (2001).
\bibitem{dna-theory}M. Muthukumar, Phys. Rev. Lett. \textbf{86}, 3188 (2001); D.K. Lubensky
and D.R. Nelson, Biophys. J. \textbf{77}, 1824 (1999).\end{thebibliography}
\end{document}